\documentclass[aps,prl,twocolumn,showpacs,floatfix,preprintnumbers,nofootinbib,superscriptaddress]{revtex4} 

\usepackage{graphicx}
\usepackage{color}
\usepackage{natbib}
\usepackage{multirow}
\usepackage{amsmath}
\usepackage[amssymb]{SIunits}
\usepackage{epstopdf}
\usepackage[draft,silent,inline,nomargin]{fixme}
\usepackage{afterpage}

\newcommand{\neff}{N_{\textrm{eff}}}

\newcommand{\fint}{f_{\textrm{int}}}
\newcommand{\ms}{m_{\nu{\textrm{,s}}}}
\newcommand{\lmeff}{\Lambda+1\nu_{\textrm{s}}+\ms}
\newcommand{\lmnu}{\Lambda+0.06{\textrm{ eV }}{\nu_{\textrm{a}}}}
\newcommand{\lnus}{\Lambda+1{\textrm{ eV }}{\nu_{\textrm{s}}}}

\begin{document}

\title{Sterile neutrinos with pseudoscalar self-interactions and cosmology}

\author{Maria Archidiacono}
\affiliation{Department of Physics and Astronomy,
 Aarhus University, 8000 Aarhus C, Denmark}

\author{Steen Hannestad}
\affiliation{Department of Physics and Astronomy,
 Aarhus University, 8000 Aarhus C, Denmark}

\author{Rasmus Sloth Hansen}
\affiliation{Department of Physics and Astronomy,
 Aarhus University, 8000 Aarhus C, Denmark}

\author{Thomas Tram}
\affiliation{Institute of Gravitation and Cosmology, University of Portsmouth, Dennis Sciama Building, Burnaby Road, Portsmouth, PO1 3FX, United Kingdom}

\begin{abstract}

Sterile neutrinos in the electronvolt mass range are hinted at by a number of terrestrial neutrino experiments. However, such neutrinos are highly incompatible with data from the Cosmic Microwave Background and large scale structure.
This paper discusses how charging sterile neutrinos under a new pseudoscalar interaction can reconcile eV sterile neutrinos with terrestrial neutrino data. We show that this model can reconcile eV sterile neutrinos in cosmology, providing a fit to all available data which is way better than the standard $\Lambda$CDM model with one additional fully thermalized sterile neutrino. In particular it also prefers a value of the Hubble parameter much closer to the locally measured value.

\end{abstract}

\pacs{14.60.St, 14.60.Pq, 98.80.Es, 98.80.Cq}

\maketitle

\section{Introduction} 

Data from the Planck satellite \cite{Planck:2015xua} and other cosmological probes have provided astoundingly precise information on cosmology. 
While some cosmological parameters, like the spatial curvature, are probed directly by the CMB measurements, others can only be inferred indirectly from a global fit to all cosmological parameters simultaneously. The prime example in the last category is the Hubble parameter which formally has been determined very accurately from Planck data \cite{Planck:2015xua}. However, the inferred value can only be trusted within the standard $\Lambda$CDM cosmological model. Furthermore there is a pronounced discrepancy between the value inferred from CMB data and the value measured directly using standard candles in the nearby universe \cite{Riess:2011yx}.
This discrepancy is mitigated if the model includes additional relativistic species (i.e. a non-standard value of $\neff$).

Concerning neutrinos, Big Bang Nucleosythesis (BBN), CMB and Large Scale Structure (LSS) measurements disfavor a fourth (sterile) neutrino with a mass in the eV range, as indicated by oscillation experiments \cite{Kopp:2013vaa,Giunti:2013waa,Gariazzo:2015rra}. There are two reasons: On the one hand one additional degree of freedom is excluded by BBN and CMB observations; on the other hand the eV mass scale is ruled out with high significance by LSS bounds. Recently, various models of neutrino self-interactions
\cite{Hannestad:2013ana,Dasgupta:2013zpn,Mirizzi:2014ama,Saviano:2014esa,Forastieri:2015paa,Chu:2015ipa}
have been proposed to reconcile sterile neutrinos in cosmology, by preventing additional neutrinos from being fully thermalized in the early Universe. 
Among the secret interaction models, the pseudoscalar model \cite{us}, where interactions are confined to the sterile sector and mediated by a light pseudoscalar, can naturally accommodate one eV-sterile neutrino in cosmology, solving the tension with BBN and CMB measurements.

\section{Model framework and production of sterile neutrinos}

Specifically, the model considered couples the mainly sterile 4th neutrino mass state, $\nu_4$, to a new light pseudoscalar, $\phi$, with mass $m_\phi \ll 1\electronvolt$ via
\begin{equation}
{\cal L} \sim g_s \phi \bar\nu_4 \gamma_5 \nu_4.
\end{equation}
This new interaction provides a background matter potential for neutrinos which is quadratic in $g_s$ and, 
provided that the coupling constant $g_{\textrm{s}}$ is larger than $10^{-6}$, the sterile neutrino production in the early Universe is significantly suppressed until after the decoupling of active neutrinos. This means that the total energy density in active neutrinos, quantified in terms of 
\begin{equation}
\neff \equiv \frac{\rho_{\nu_a}+\rho_{\nu_s}}{\rho_0}
\end{equation}
with $\rho_0 = (4/11)^{4/3} \rho_\gamma$, can be significantly lower than the $\neff \sim 4$ predicted in the absence of non-standard interactions.

However, even in the presence of a significant non-standard matter potential, sterile and active neutrinos can still equilibrate at later times, and if the sterile neutrino has a mass of order 1 eV this can lead to conflict with the cosmological bound on neutrino mass \cite{Saviano:2014esa}. 

Here we will refrain from a detailed calculation of how efficient the equilibration is and simply use the most conservative assumption possible.
Naively we expect that full equilibration leads to $\neff/4$ in the sterile sector and $3 \neff/4$ in the active sector. However, the pseudoscalar interaction leads to a strong coupling between sterile neutrinos and pseudoscalars at late times. If this happens prior to active-sterile equilibration the division will be $11 \neff/32$ in the strongly coupled $\nu_s-\phi$ fluid and $21 \neff/32$ in the remaining non-interacting and massless neutrinos. As we will see in the next section structure formation data prefer a smaller fraction of the strongly interacting component, so taking the fraction to be $11/32$ is the most conservative choice.

\section{Late time phenomenology and fit to structure formation data}

The late-time phenomenology of the pseudoscalar model leads to a collisional recoupling of sterile neutrinos via the process $\nu_s \nu_s \leftrightarrow \phi \phi$ before their non-relativistic transition (assuming $g_{\textrm{s}}>10^{-6}$ and $\ms>1$ eV) and before recombination. In the collisional regime, neutrinos and pseudoscalars are not free-streaming, but rather behave as a single fluid with no anisotropic stress.
The impact on the CMB temperature power spectrum consists of an enhancement of the monopole term \cite{Archidiacono:2013dua}, which would spoil the Planck measurements if the collisional regime was extended to all neutrino species, including active neutrinos.

\begin{figure}
\begin{center}
\includegraphics[width=\columnwidth]{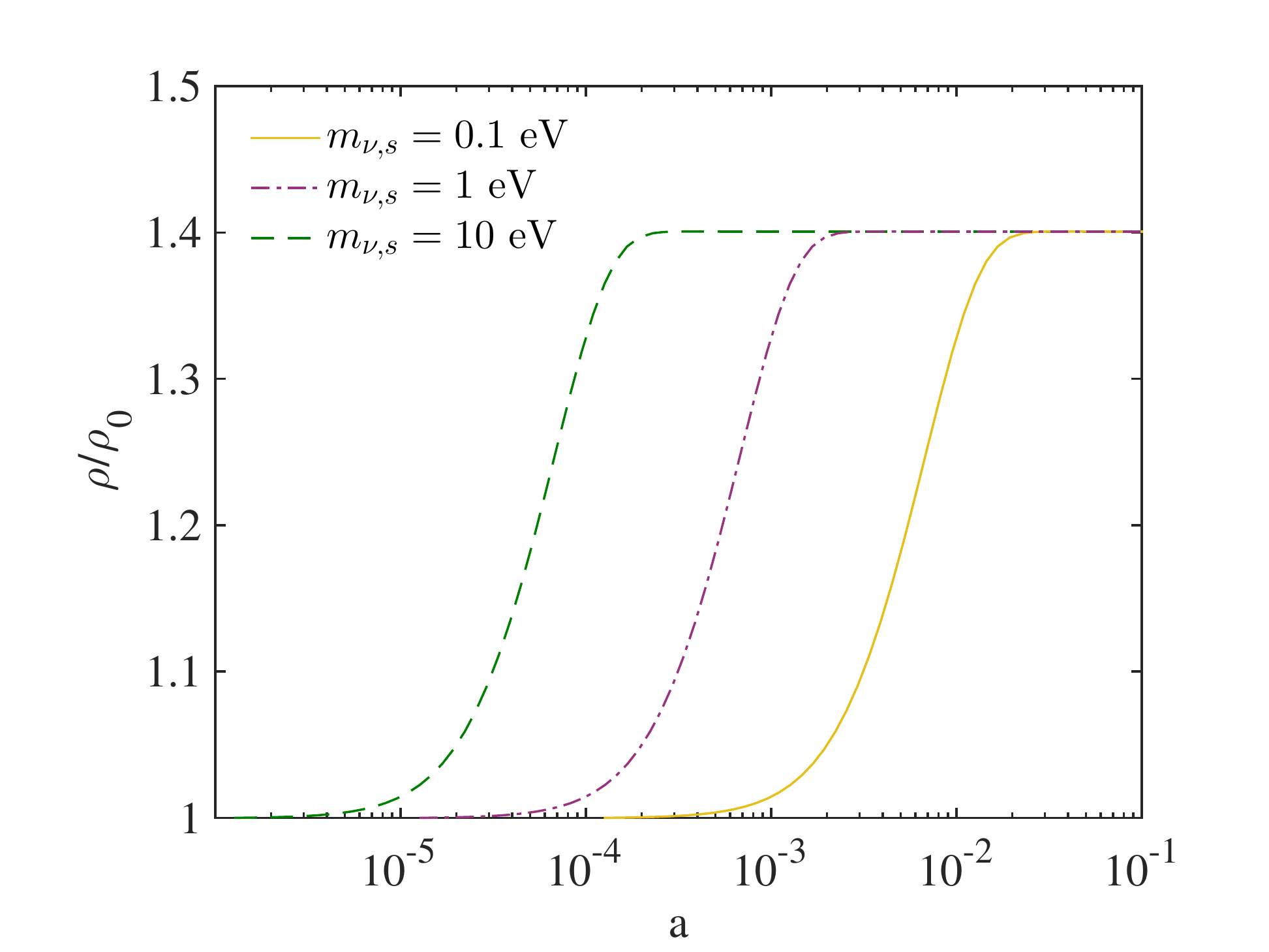}\\
\includegraphics[width=\columnwidth]{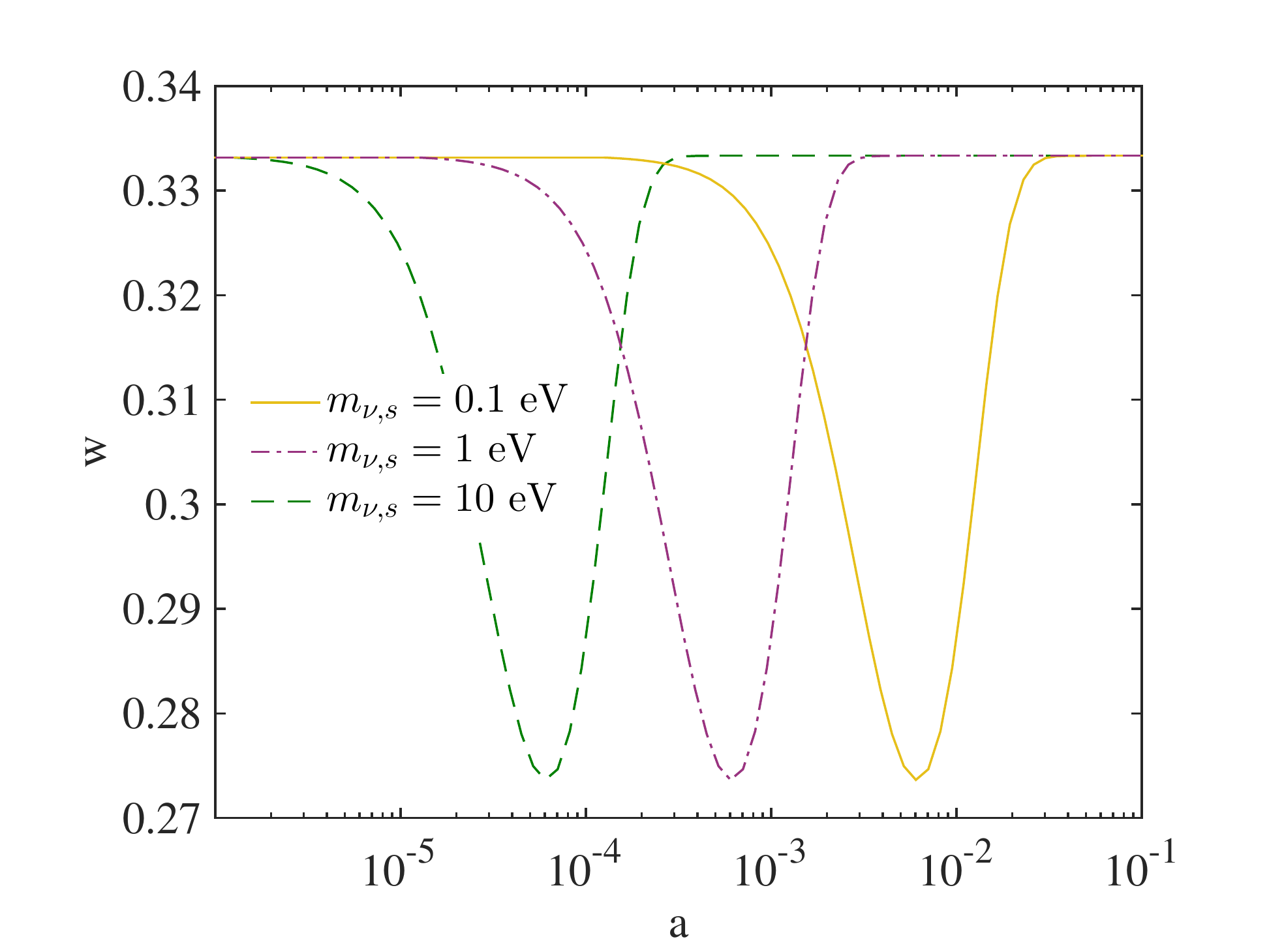}
\end{center}
\caption{Upper panel: Relative increase in the pseudo scalar-sterile neutrino energy density compared to the energy density of one active neutrino, due to sterile neutrino annihilations. Lower panel: Temporary suppression of the pseudo scalar-sterile neutrino equation of state parameter.
}
\label{fig:rhow}
\end{figure}

At temperatures higher than the sterile neutrino mass the combined fluid is fully relativistic with an effective equation of state parameter of $w=1/3$.
However, when sterile neutrinos go non relativistic, they annihilate into pseudoscalars $\nu\nu \rightarrow \phi\phi$, while the inverse process $\phi\phi \rightarrow \nu\nu$ becomes kinematically prohibited. This increases the energy density of the combined fluid relative to that of a fully relativistic fluid.
During this process the pressure of the combined fluid also drops relative to energy density because of the importance of the rest mass of the sterile neutrino \cite{Hannestad:2000gt,Beacom:2004yd}.
However, for much lower temperatures the fluid consists only of pseudoscalars and is again fully relativistic with $w=1/3$.
The redshift where deviations in $\rho$ and $w$ set in depends directly on the sterile neutrino mass. We have shown the redshift evolution of $\rho$ and $w$  in the upper and lower panel of Figure \ref{fig:rhow}, respectively. 

At the end of the annihilation processes the cosmic abundance of sterile neutrinos is highly suppressed and there is little suppression of matter fluctuations on small scales. This is in stark contrast to the $\Lambda$CDM model where the presence of a free streaming massive sterile neutrino causes a dramatic suppression of the matter power spectrum.

\begin{figure}
\begin{center}
\includegraphics[width=\columnwidth]{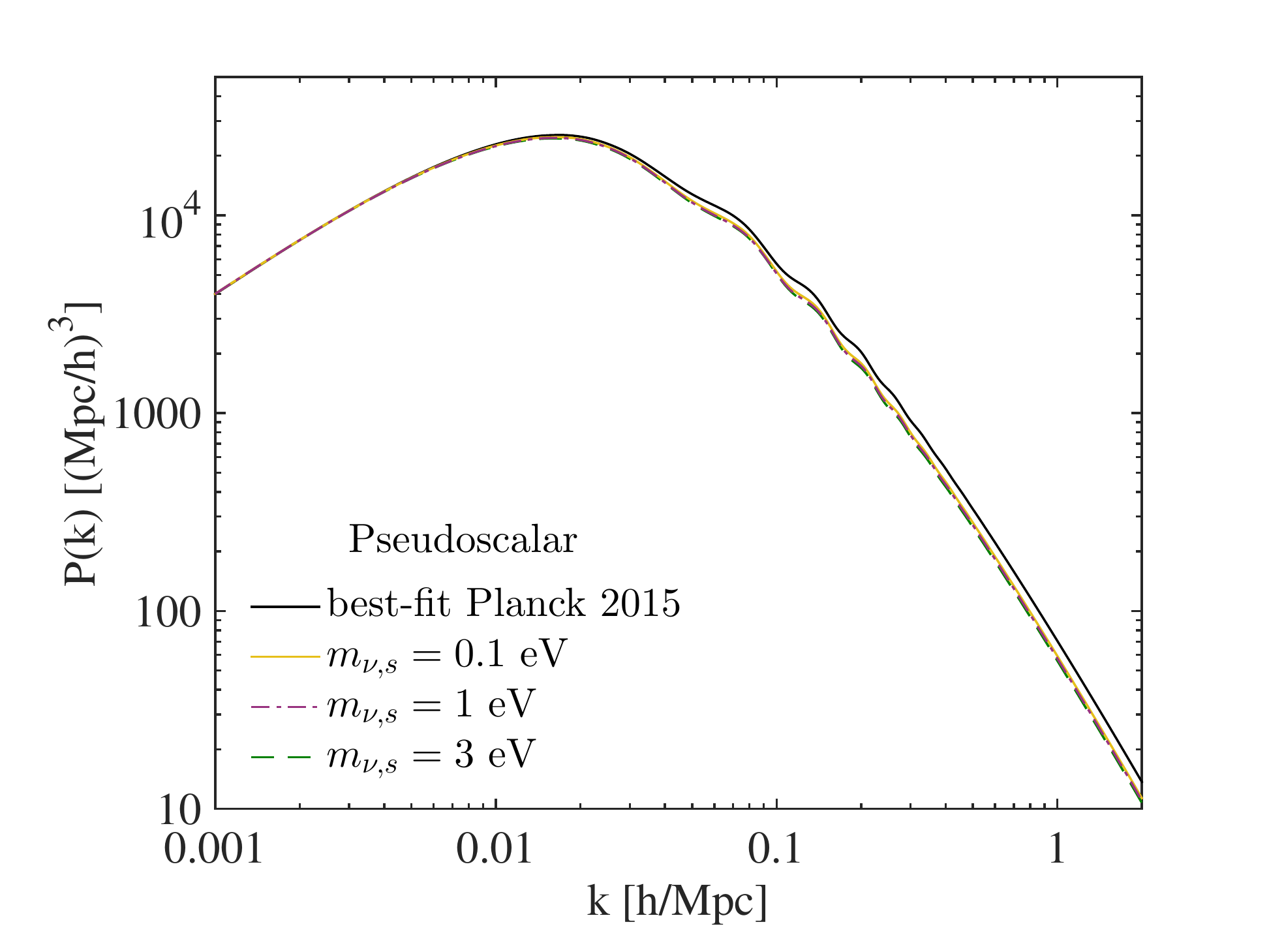}
\includegraphics[width=\columnwidth]{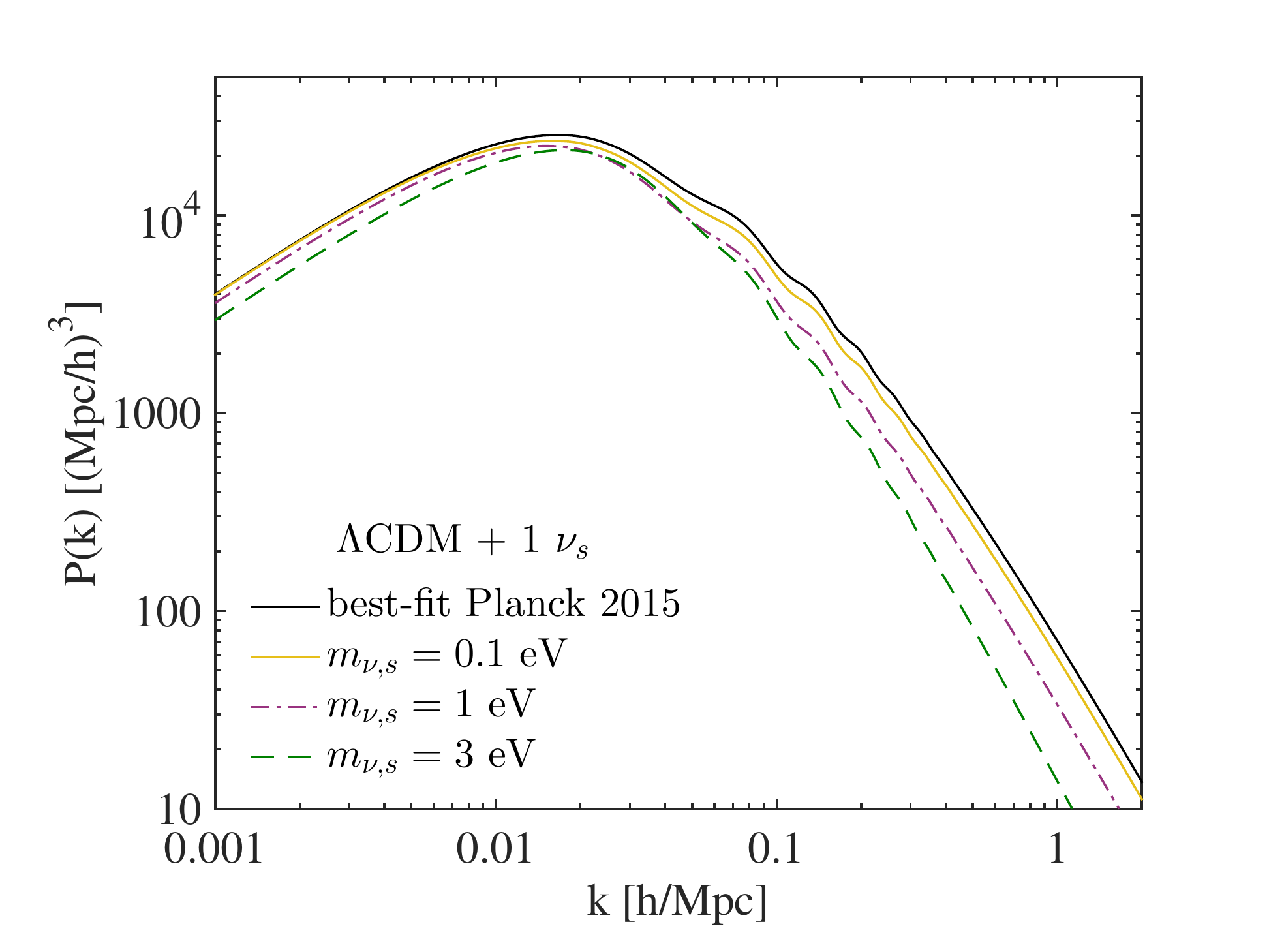}
\end{center}
\caption{Linear matter power spectra for $\lmeff$ (lower panel) and the pseudoscalar model (upper panel) for various values of the sterile neutrino mass and for $\neff=4.046$ (in the pseudoscalar model $11/32 \times 4.046$ are strongly interacting).
For comparision, the matter power spectrum obtained with Planck 2015 best-fit is also shown.
}
\label{fig:mpk}
\end{figure}

The difference between the two scenarios can easily be seen in Figure~\ref{fig:mpk}
where we show the matter power spectrum obtained for three different sterile neutrino masses: in the pseudoscalar scenario the increase in the sterile neutrino mass does not lead to the suppression of power at small scales that one finds in the $\Lambda$CDM model.

\subsection{$\Lambda$CDM with massive sterile neutrinos}

If the only non-standard physics is the addition of a sterile neutrino with the mass and mixing needed to explain the short baseline data, the expectation is that the additional species are almost fully thermalised in the early universe (for a recent treatment see e.g.\ \cite{Hannestad:2015tea}). The prediction therefore is that $\neff \sim 4$ and $m_s \sim 1$ or $\sim 3$ eV \cite{Kopp:2013vaa,Giunti:2013waa}. Thus, when testing the $\Lambda$CDM model with a varying sterile neutrino mass $\ms$, we will keep $\neff=4.046$. Hereafter we will refer to this model as the $\lmeff$ model, while the pure $\Lambda$CDM model with 3.046 massless active neutrinos is denoted by $\Lambda$ and the model with $1$ eV fully thermalized sterile neutrino is denoted by $\lnus$.

\subsection{Pseudoscalar model}

In the pseudoscalar model $\neff$ is a free parameter depending on $g_s$. However, unless the pseudoscalar is thermalised by other means in the early Universe the expectation is that $3 \lesssim \neff \lesssim 4$ with $\neff \sim 3$ corresponding to $g_s \gtrsim 10^{-5}$ and $\neff \sim 4$ corresponding to $g_s \lesssim 10^{-6}$.

Assuming that actives and steriles completely equilibrate via oscillations prior to the timescales of interest to CMB and large scale structure, the expection is that 11/32 of the total energy density in all neutrinos and the pseudoscalar is in the $\nu_s-\phi$ fluid. 
However, this ratio could be smaller if equilibration is incomplete or if steriles and pseudoscalars equilibrate after active-sterile equilibration.
In most runs we fix the ratio, $\fint$, to 11/32, but we have also tested the case where $\fint$ is allowed to vary.

We stress that each value of $g_s$ corresponds to \emph{one} value of $\neff$ and \emph{one} value of $\fint$, but while $\neff$ can be found as a function of $g_s$~\cite{us}, the calculation yielding $\fint$ is unfeasible to perform using current technology, and we simply leave $\fint$ as a free parameter when considering partial equilibration. When also letting $\neff$ vary freely, the result is a two-dimensional parameter space, where the pseudoscalar models trace out a one-dimensional path.

\subsection{Comparison with data}

In order to test the ability of the pseudoscalar model to fit cosmological data and to constrain its parameters, we perform a Markov Chain Monte Carlo analysis using \texttt{CosmoMC} \cite{Lewis:2002ah} with a modified version of the Boltzman solver \texttt{CAMB} \cite{Lewis:1999bs}. Our basic data set is based on Planck 2015 high multipole temperature data and low multipole polarization data (PlanckTT+lowP), implemented according to the prescription of Ref. \cite{Aghanim:2015xee}. Additional data sets are the Hubble Space Telescope (HST) prior on the Hubble constant from direct measurements \cite{Riess:2011yx} and Baryonic Acoustic Oscillations (BAO), including 6dFGS \cite{Beutler:2011hx}, SDSS-MGS \cite{Ross:2014qpa}, BOSS-LOWZ BAO \cite{Anderson:2012sa} and CMASS-DR11 \cite{Anderson:2013zyy}.

{\it $\chi^2$ results ---}

In order to get a feeling for how well the models fit the data compared with each other we have obtained best-fit $\chi^2$ values for a variety of different cases (see Table \ref{tab:chi}).
Before comparing values for the different models, we should state clearly the question we want to ask. We are in fact less interested in comparing the pseudoscalar model with the standard $\Lambda$CDM model. Instead, we would like to compare these two models in a possible future where terrestrial experiments have confirmed the existence of an eV-scale sterile neutrino. So the two models that we should compare are the pseudoscalar model and the $\Lambda$CDM model with an additional fully thermalised neutrino of mass around 1 eV or higher.

\begin{table*}[p]
\begin{center}
\begin{tabular}{lccccl}
\hline
\hline
Data & $\chi^2_{\rm tot}$ & $\chi^2_{\rm CMB}$ & $\chi^2_{\rm HST}$ & $\chi^2_{\rm BAO}$ & Model\\[0.1cm]
\hline
 CMB & $11265.8$ & $11265.8$ & $--$ & $--$ &  P \\ [0.1cm]
 & $11259.1$ & $11259.1$ & $--$ & $--$ &  P + $\fint$ \\ [0.1cm]
 & $11274.8$ & $11274.8$ & $--$ & $--$ &  $\lnus$ \\ [0.1cm]
 & $11260.3$ & $11260.3$ & $--$ & $--$ &  $\Lambda$ \\ [0.1cm]
\hline
 CMB+HST & $11265.8$ & $11265.8$ & $0.0$ & $--$ & P \\ [0.1cm]
 & $11260.2$ & $11260.0$ & $0.2$ & $--$ &  P + $\fint$ \\ [0.1cm]
 & $11279.2$ & $11275.5$ & $3.7$ & $--$ &   $\lnus$\\ [0.1cm]
 & $11266.1$ & $11262.6$ & $3.5$ & $--$ &   $\Lambda$\\ [0.1cm]
\hline
CMB+BAO & $11270.5$ & $11266.3$ & $--$ & $4.2$ &   P \\ [0.1cm]
 & $11263.8$ & $11259.4$ & $--$ & $4.4$ &   P + $\fint$ \\ [0.1cm]
 & $11288.1$ & $11279.6$ & $--$ & $8.5$ &  $\lnus$\\ [0.1cm]
 & $11264.7$ & $11260.2$ & $--$ & $4.5$ &  $\Lambda$\\ [0.1cm]
\hline
All & $11272.9$ & $11267.3$ & $0.7$ & $4.9$ &   P \\ [0.1cm]
 & $11266.0$ & $11259.6$ & $1.8$ & $4.6$ &  P + $\fint$ \\ [0.1cm]
 & $11288.7$ & $11280.2$ & $1.7$ & $6.8$ &  $\lnus$\\ [0.1cm]
 & $11270.8$ & $11260.6$ & $5.8$ & $4.4$ &  $\Lambda$ \\ [0.1cm]
\hline
\hline
\end{tabular}
\end{center}
\caption{Best-fit $\chi^2$ for various models and various data set combinations. The labels for the various models follow the prescription described in the text. Data-set combinations are labeled according to the text. The ``All'' case refers to CMB+HST+BAO. The best-fit $\chi^2$ values were obtained with the \texttt{BOBYQA} routine implemented in \texttt{CosmoMC}. The obtained values of $\chi^2$ are typically within $\Delta \chi^2 \sim 1$ of the true global best-fit value. 
}
\label{tab:chi}
\end{table*}

\begin{table*}[p]
\begin{center}
\begin{tabular}{lcccccl}
\hline
\hline
 Data &$n_s$&$\neff$ & $\ms {\textrm{[eV]}}$ & $H_0 {\textrm{[km/s/Mpc]}}$ & $\fint$ & Model\\[0.1cm]
\hline
CMB  &$0.978^{+0.014}_{-0.012}$& $3.66^{+0.28}_{-0.36}$ & $3.05^{+1.1}_{-0.76}$ & $74.0^{+2.3}_{-3.0}$ & $11/32$ & P\\[0.1cm]
  &$0.9779^{+0.0080}_{-0.011}$& $3.42^{+0.11}_{-0.36}$ & $5.7^{+1.7}_{-2.0}$ & $71.2^{+1.5}_{-2.6}$ & $0.111^{+0.050}_{-0.062}$ & P + $\fint$ \\[0.1cm]
  &$0.9994\pm 0.0052$& $4.046$ & $<0.364$ & $74.5^{+1.6}_{-0.86}$ & $--$ & $\lmeff$\\[0.1cm]

\hline
CMB+HST &$0.977^{+0.012}_{-0.0079}$& $3.64\pm 0.19$ & $3.00^{+1.2}_{-0.67}$ & $73.8\pm 1.4$ & $11/32$ & P\\[0.1cm]
  &$0.9848\pm 0.0074$& $3.59\pm 0.19$ & $5.3^{+1.4}_{-1.8}$ & $72.8\pm 1.4$ & $0.124^{+0.054}_{-0.060}$ & P + $\fint$ \\[0.1cm]
  &$0.9990\pm 0.0045$& $4.046$ & $<0.284$ & $74.4^{+2.0}_{-2.2}$ & $--$ & $\lmeff$\\[0.1cm]

\hline
CMB+BAO &$0.968^{+0.011}_{-0.0057}$& $3.34^{+0.11}_{-0.25}$ & $3.61^{+0.97}_{-0.43}$ & $70.75^{+0.94}_{-1.4}$ & $11/32$ & P\\[0.1cm]
  &$0.9706^{+0.0052}_{-0.0059}$& $<3.55$ & $6.0\pm 1.9$ & $69.55^{+0.76}_{-1.2}$ & $0.104^{+0.048}_{-0.060}$ & P + $\fint$ \\[0.1cm]
 &$0.9945\pm 0.0038$& $4.046$ & $0.26^{+0.10}_{-0.13}$ & $72.38\pm 0.61$ & $--$ & $\lmeff$\\[0.1cm]

\hline
All &$0.9737^{+0.0075}_{-0.0057}$& $3.49\pm 0.18$ & $3.77^{+0.64}_{-0.51}$ & $71.8\pm 1.0$ & $11/32$ & P\\[0.1cm]
 &$0.9762\pm 0.0058$& $3.45^{+0.16}_{-0.18}$ & $5.7^{+1.2}_{-1.7}$ & $70.9\pm 1.0$ & $0.125^{+0.051}_{-0.057}$ & P + $\fint$ \\[0.1cm]
 &$0.9946\pm 0.0038$& $4.046$ & $0.24^{+0.10}_{-0.12}$ & $72.54\pm 0.58$ & $--$ & $\lmeff$\\[0.1cm]

\hline
\hline
\end{tabular}
\end{center}
\caption{Marginalized constraints are given at $1\sigma$, while upper bounds are given at $2\sigma$ for the pseudoscalar model (with and without varying $\fint$) and for the $\Lambda$CDM model with one additional fully thermalized massive neutrino. Data-set combinations are the same as in Table~\ref{tab:chi}}.
\label{tab:marg}
\end{table*}

\begin{figure}
\begin{center}
\includegraphics[width=\columnwidth]{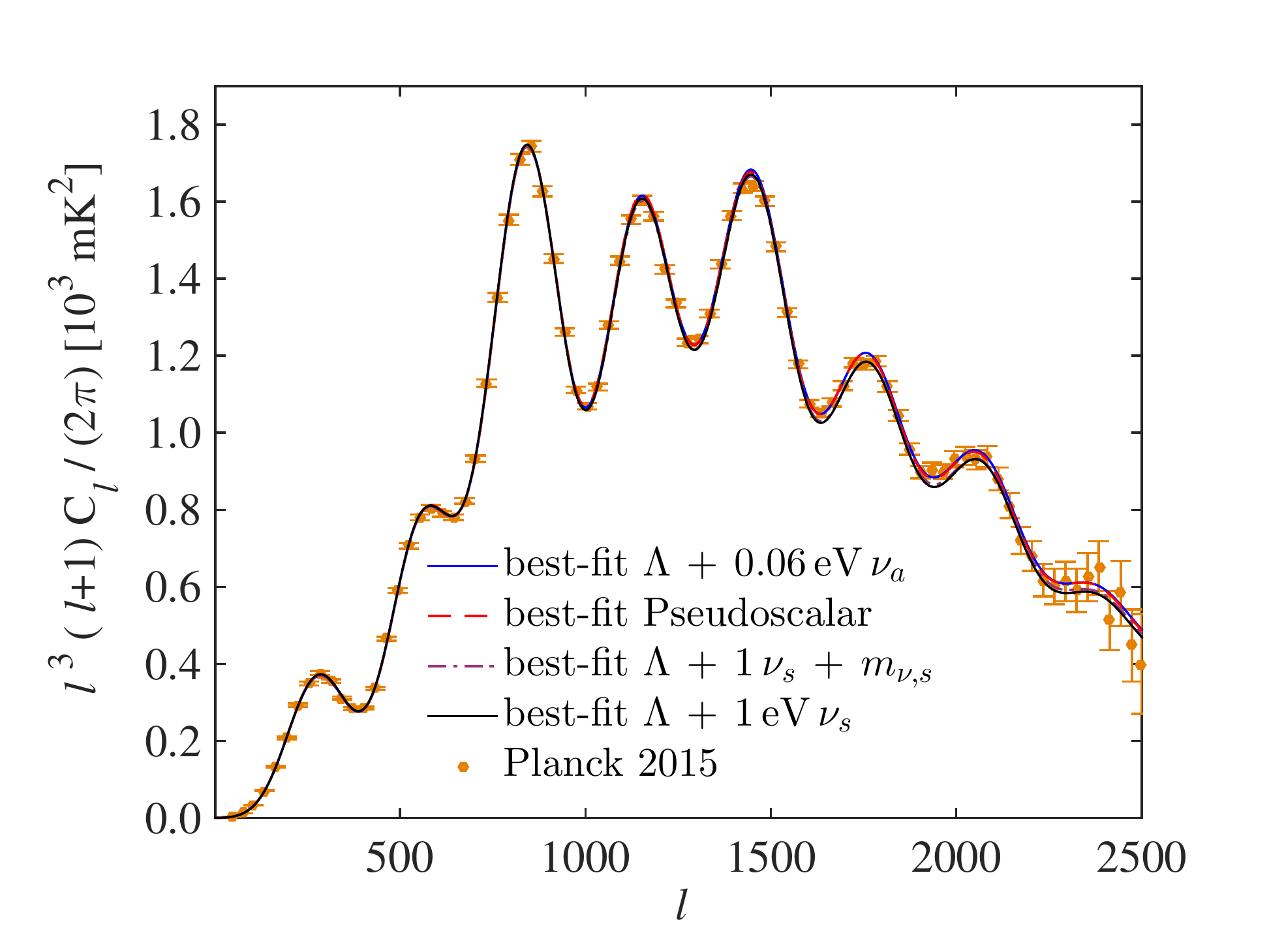}\\
\includegraphics[width=\columnwidth]{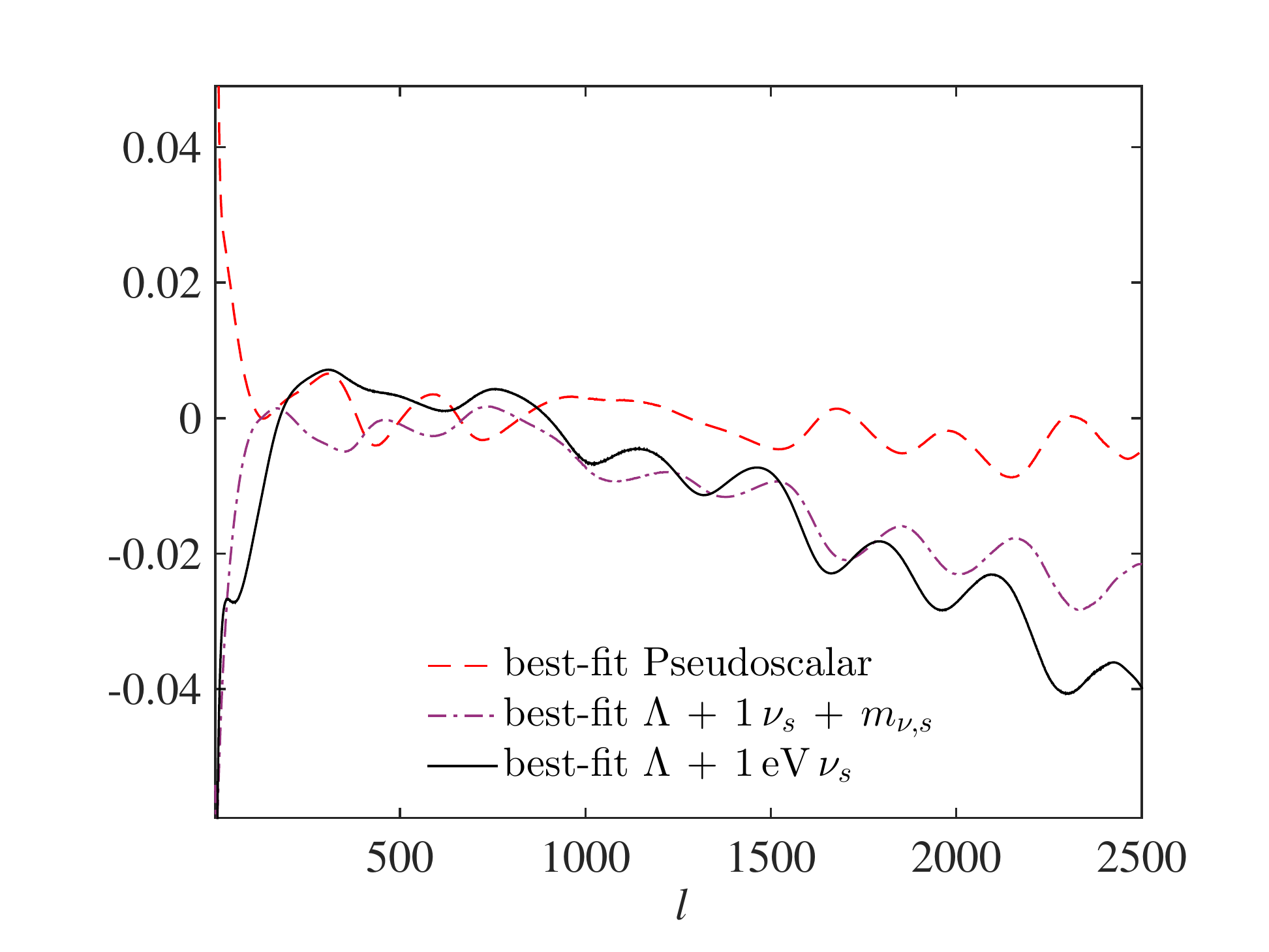}\\
\end{center}
\caption{Upper panel: Temperature anisotropies power spectra for various models: Pseudoscalar (red/dashed line), $\lmnu$ (blue/solid line), $\lmeff$ (purple/dot-dashed line), $\lnus$ (black/solid line). Lower panel: Relative errors for the models listed above compared to the $\lmnu$ model. 
In all cases the spectra plotted are with the best-fit parameters obtained when fitting to CMB data.}
\label{fig:clsl3}
\end{figure}

If only CMB data is used the pseudoscalar model with fixed $\fint$ has a slightly higher $\chi^2$ than $\Lambda$CDM and the model with varying $\fint$ has a slightly lower $\chi^2$. Even $\Lambda$CDM with a 1 eV sterile neutrino does not have a significantly higher $\chi^2$. However, especially when BAO data is included, the sensitivity to the neutrino rest mass increases drastically and $\chi^2$ for $\Lambda$CDM with a 1 eV sterile neutrino becomes much worse - formally excluding it at approximately $5\sigma$ relative to pure $\Lambda$CDM. However, the pseudoscalar model again has a $\chi^2$ comparable to $\Lambda$CDM and thus provides a significantly better fit than $\Lambda$CDM  with a 1 eV sterile neutrino, as shown in Figure \ref{fig:clsl3}. In conclusion, in the absence of evidence for eV sterile neutrinos from short baseline data, cosmological data does not merit the inclusion of sterile neutrinos or sterile neutrino interactions. However, if the existence of eV sterile neutrinos is confirmed by future short baseline data, new physics is needed in order to reconcile them with cosmology. Our analysis shows that the pseudoscalar interaction drastically reduces $\chi^2$ and allows for a fit as good as pure $\Lambda$CDM.

{\it Parameter constraints ---}

In Table \ref{tab:marg} the $1\sigma$ marginalized constraints or the $2\sigma$ limits on $n_s$, $\neff$, $\ms$, $H_0$ are reported for various data set combinations and for different models.

From the analysis it is clear that although $\neff=4$ is marginally allowed in the pseudoscalar model (within $2\sigma$ when fitting only CMB data), a value below 4 is preferred. This indicates that relatively high values of $g_s$ are favored and sterile neutrinos are partially thermalized in the early universe. E.g., when BAO are included in the analysis and $\fint$ is allowed to vary, $\neff<3.55$ at 95\% c.l. This upper bound on $\neff$ translates into a lower bound on the coupling $g_s>3\times10^{-6}$, which is consistent with the expectations.

The preference for $\fint>0$ at more than $2\sigma$ underlines that additional neutrino species cannot be free-streaming at recoupling, thus, if the short baseline experiments will confirm the existence of a sterile neutrino, non standard interactions will be required in order to accommodate a fourth neutrino in cosmology.

Note that the usual correlation between the scalar spectral index $n_s$ and $\neff$ is driven by diffusion damping at large $l$ which are absent for a tightly coupled fluid. This effect is clearly seen on the lower panel of Figure~\ref{fig:clsl3}. That means that the value of $n_s$ is virtually unchanged with respect to standard $\Lambda$CDM while the $\lnus$ model prefers a much higher value as shown in Table~\ref{tab:marg} .

\begin{figure}
\begin{center}
\includegraphics[width=\columnwidth]{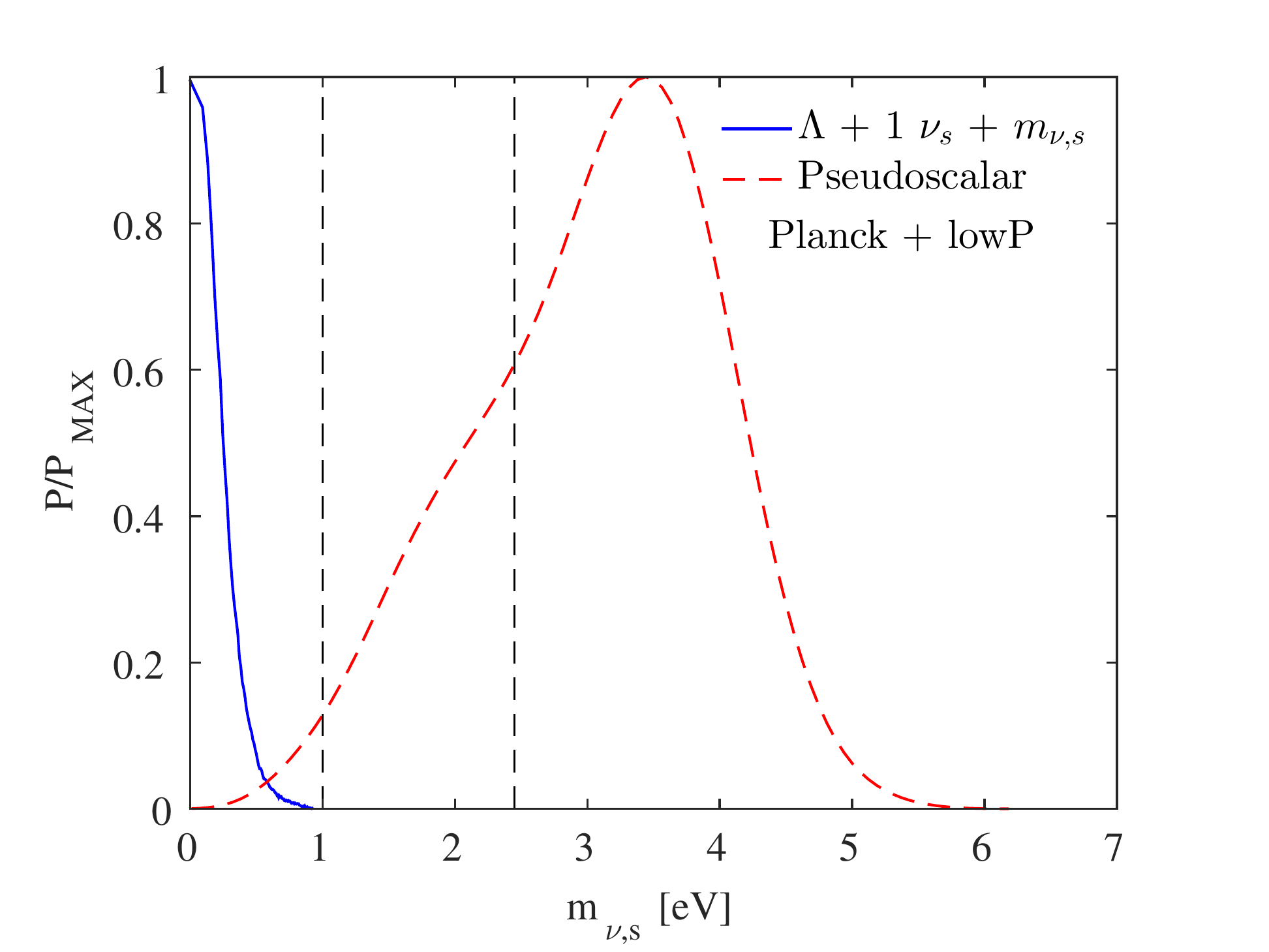}
\end{center}
\caption{One dimensional posterior for $\ms$ in the $\lmeff$ model and in the pseudoscalar scenario for the combination of data PlanckTT+lowP. The vertical lines show the best fit for the sterile neutrino masses obtained by the global oscillation data analysis in the 3+1 scenario \cite{Kopp:2013vaa}
}
\label{fig:mnus}
\end{figure}

Considering the sterile neutrino mass, when only CMB data are considered, $\ms<0.364$ eV at $2\sigma$, this upper limit further tightens when BAO are included; the eV range is excluded at high significance in a pure $\Lambda$CDM model with free streaming neutrinos.
However, the late time phenomenology of the pseudoscalar model makes sterile neutrinos fully consistent with the eV-mass range: the overlap with the results of the global fit in the 3+1 scenario \cite{Kopp:2013vaa} is not only in the region around $\Delta m^2 \sim 1 \, {\rm eV}^2$ but also with $\Delta m^2 \sim 6 \, {\rm eV}^2$ (see Figure~\ref{fig:mnus}).

\begin{figure}
\begin{center}
\includegraphics[width=\columnwidth]{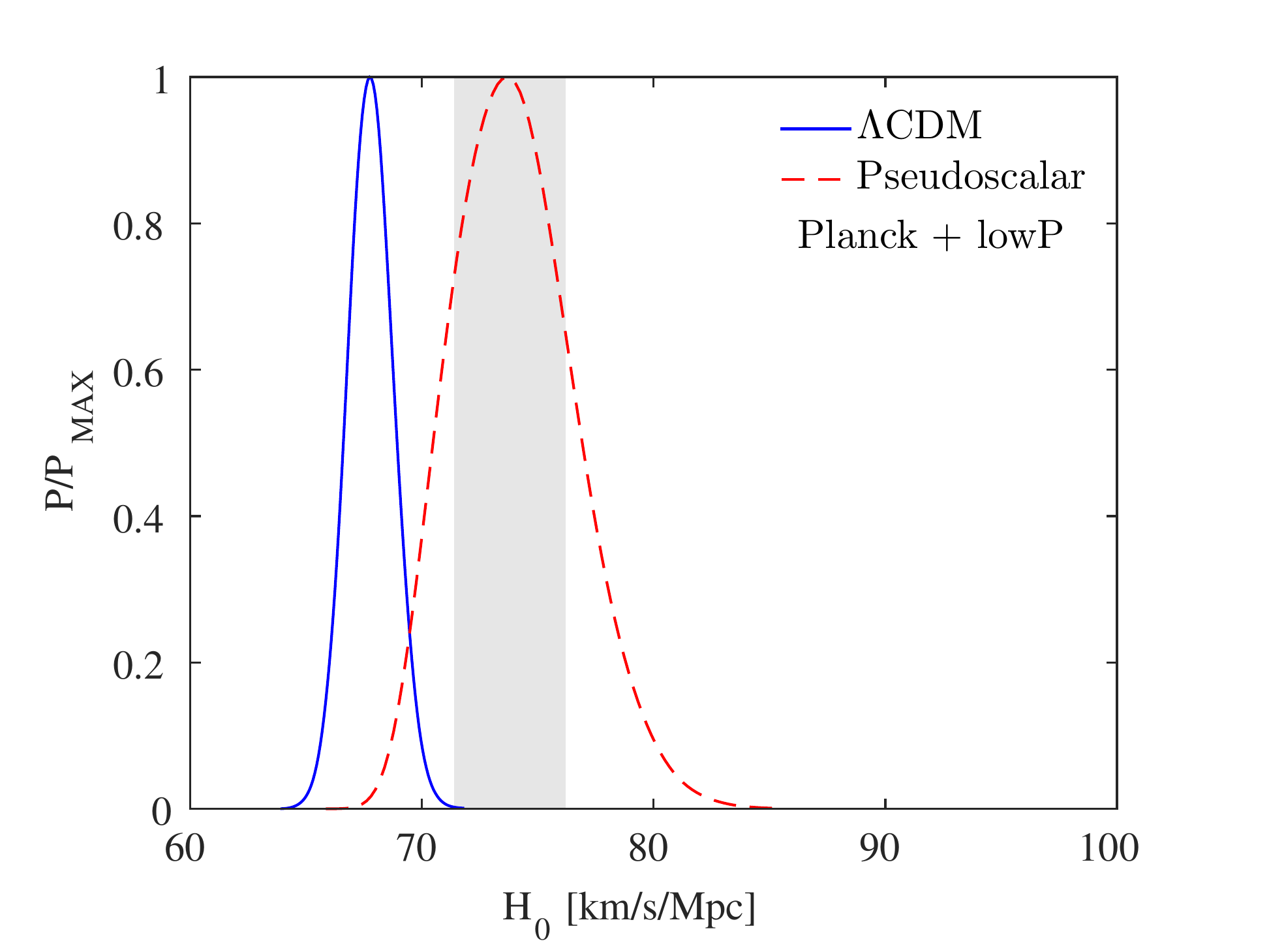}
\end{center}
\caption{One dimensional posterior for $H_0$ in the $\Lambda$CDM model and in the pseudoscalar scenario for the combination of PlanckTT+lowP data. The grey region shows the $1\sigma$ confidence interval from direct measurements.
}
\label{fig:h0}
\end{figure}

In Figure \ref{fig:h0} we show that models with a strongly self-interacting dark radiation component provide a remarkably good fit to CMB data with a preferred value of the Hubble parameter fully consistent with local measurements.

\begin{figure*}
\begin{center}
\includegraphics[width=2\columnwidth]{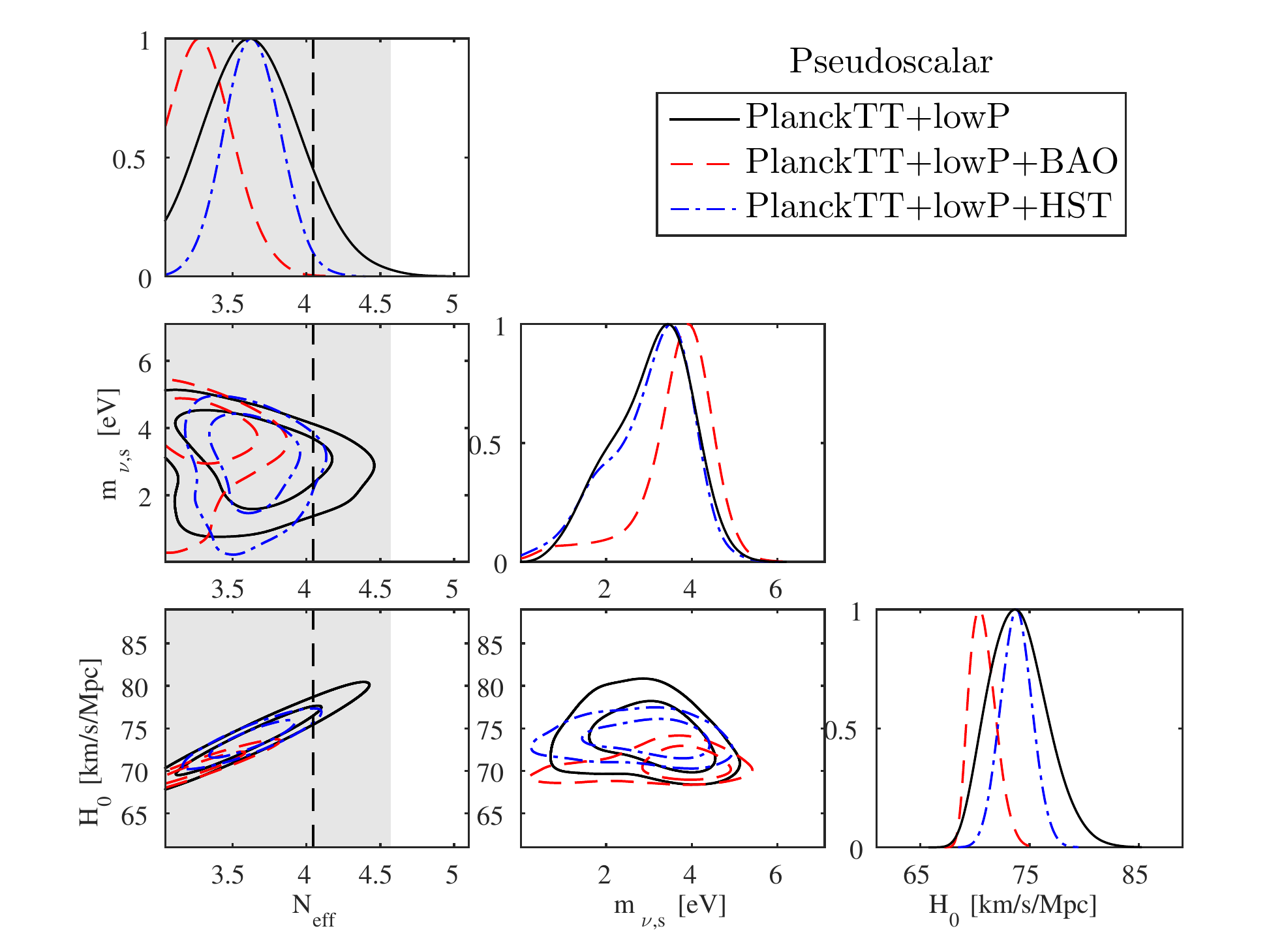}
\end{center}
\caption{Triangle plot in the parameter space $(\neff, \, \ms, \, H_0)$ showing the 1D marginalized posteriors and the 2D marginalized contours obtained with various data set combinations in the pseudoscalar scenario. The partial thermalization of pseudoscalars and sterile neutrinos in the early Universe can make one sterile neutrino consistent with a value of $\neff$ between $3$ and $4.57$ (grey shaded region), depending on $g_s$; while the $\Lambda$CDM model has to be consistent with one fully thermalized additional degree of freedom ($\neff = 4.046$, black/dotted vertical line) in order to account for one sterile neutrino with large mixing angle.
}
\label{fig:tri}
\end{figure*}

Finally, Figure \ref{fig:tri} shows one-dimensional posteriors and $1$ and $2$ $\sigma$ marginalized contours for $(\neff, \, \ms, \, H_0)$, obtained within the pseudoscalar scenario and with various data set combinations.

\section{Discussion}

We have tested the pseudoscalar model against the most precise available cosmological data and found that the model is generally compatible with data, providing at least as good a fit as the standard $\Lambda$CDM model. Furthermore the fit is vastly better than $\Lambda$CDM with an additional sterile neutrino in the eV mass range.

If the eV sterile neutrino interpretation of short baseline data turns out to be true cosmology is faced with a very serious challenge. Taken at face value such a model is excluded by CMB and large scale structure data at least at the 5$\sigma$ level.
With this in mind it is clear that accommodating eV sterile neutrinos requires addition of new physics either in cosmology or in the neutrino sector (see e.g.\ \cite{Hamann:2011ge} for a discussion). 

The model discussed here provides a simple and elegant way of reconciling eV sterile neutrinos with precision cosmology. 
We again stress that this model has a late-time phenomenology very different from models with purely free-streaming neutrinos and that it could well be possible to test details of the model with the greatly enhanced precision of future cosmological surveys such as Euclid~\cite{Laureijs:2011gra}.

Finally, it is interesting that a recent study by Lesgourgues {\it et al.} \cite{Lesgourgues:2015wza} find that current cosmological data prefers relatively strong self-interactions between dark matter and a new dark radiation component. While the model presented here cannot provide such dark matter interactions at the required strength unless the fundamental coupling becomes close to unity, it could be a another indication that we are seeing the first signs of new, hidden interactions in the dark sector.

\acknowledgments{
We thank Jan Hamann for comments on the manuscript. STH and TT thank the Mainz Institute for Theoretical Physics for its hospitality and support during the completion of this work.
MA acknowledges partial support from the European Union FP7 ITN INVISIBLES (Marie Curie Actions, PITN- GA-2011- 289442)}

\end{document}